%% file: main.tex
\documentclass[sigplan,10pt]{acmart}
\usepackage{algorithm}
\usepackage{algorithmicx}
\usepackage{algcompatible}
\usepackage{xspace}
\renewcommand\footnotetextcopyrightpermission[1]{}
\newcommand{\rev}[1]{\textcolor{black}{#1}}
\newcommand{\wy}[1]{\textcolor{black}{#1}}
\newcommand{\wyRe}[1]{\textcolor{black}{#1}}
\newcommand{\compresses}[1]{\textcolor{black}{#1}}
\newcommand{\restruct}[1]{\textcolor{black}{#1}}
\newcommand{\final}[1]{\textcolor{black}{#1}}
\newcommand{\sysname}{Libra\xspace}
\newcommand{\sysid}{libra}

\makeatletter
\renewcommand{\@titlefont}{\fontsize{18}{19}\selectfont\bfseries\sffamily}
\makeatother

\begin{document}

\title[\sysname]{\sysname: Taming Attention Workload Skew in Long-Context LLM Training with Bounded Attention Pools}

\author{Yan Wang\textsuperscript{1,*}, Xiulong Yuan\textsuperscript{2}, Kaiming Yang\textsuperscript{3}, Jiaxuan Peng\textsuperscript{2}, Pengju Lu\textsuperscript{1}, Mingzhen Li\textsuperscript{1}, Zhipeng Zhang\textsuperscript{4}, Chang Si\textsuperscript{2}, Zhixiang Ruan\textsuperscript{2}, Hongqing Chen\textsuperscript{2}, Linlang Jiang\textsuperscript{2}, Siyu Wang\textsuperscript{2}, Langshi Chen\textsuperscript{2}, \\Rui Men\textsuperscript{2}, Man Yuan\textsuperscript{2}, Guangming Tan\textsuperscript{1}, Yong Li\textsuperscript{4}, Weile Jia\textsuperscript{1}, and Jingren Zhou\textsuperscript{2}}

\affiliation{
\institution{\textsuperscript{1}University of Chinese Academy of Sciences;
\textsuperscript{2}Alibaba Group;
\textsuperscript{3}National University of Singapore;
\textsuperscript{4}Unaffiliated}
\city{}
\country{}
}
\thanks{
\textsuperscript{*}Work done during an internship at Alibaba Group.
Corresponding authors: Weile Jia \textless jiaweile@ict.ac.cn\textgreater, Mingzhen Li \textless limingzhen@ict.ac.cn\textgreater.}

\renewcommand{\shortauthors}{Yan Wang, Xiulong Yuan}

\begin{abstract}
Long-context LLM training suffers from a load-balancing problem. Packing samples into fixed-token sequences balances memory and linear-cost operators, but the dominant attention cost scales with the sum of squared sequence lengths. Consequently, equally sized packed sequences drawn from a long-tailed corpus can carry substantially different attention workloads, creating data-parallel stragglers and pipeline bubbles.
Existing approaches are either sensitive to outliers or suffer from an expanding communication domain as the data-parallel degree grows.

We present \sysname, which operationalizes the law of large numbers (LLN) as a scaling principle for load balancing in long-context LLM training.
\sysname groups packed sequences and the CP groups processing them into fixed-size \emph{sequence pools}, where corresponding attention computation forms \emph{attention pools}. As DP scales out, \sysname increases the number of pools rather than the size of each pool, thereby bounding the scope of attention exchange.
LLN explains why a moderate pool can smooth workload variation; \emph{Variance-Reduced Sequence Placement} makes this design effective for finite, long-tailed workloads.
Within each pool, \emph{Tiled Attention Pooling} dispatches 2D SH-Tiles to balance attention.
\wyRe{On three production Qwen3 models (8B, 30B, 235B) and 256K- and 1M-token production workloads, \sysname improves end-to-end training throughput over the strongest evaluated baseline (WLB-LLM) by $44\%$ on average and up to $68\%$ at 256 GPUs.}
\sysname has accumulated hundreds of thousands of GPU-hours \final{on large-scale training jobs} spanning 32K to 1M tokens while preserving training semantics.
\end{abstract}

\settopmatter{printfolios=true}
\settopmatter{printacmref=false}
\maketitle

\input{1-introduction.tex}
\input{2-background.tex}
\input{3-motivation.tex}

\input{4-design.tex}
\input{5-implementation.tex}
\input{6-evaluation.tex}
\input{7-relatedwork.tex}
\input{9-conclusion.tex}

\bibliographystyle{ACM-Reference-Format}
\bibliography{sample-base,relatedwork}

\end{document}

%% file: 1-introduction.tex
\section{Introduction}
\label{sec:introduction}

Long contexts are increasingly important to LLM applications such as coding, reasoning, and data analysis~\cite{an2024make,yu2025memagent,jin2024llm}. Training on long contexts, however, exposes an acute load-balancing problem. \final{Sample} lengths in production corpora are highly skewed~\cite{longalign2024,longlora2024,redpajama2024}, and \final{the production corpus used in this paper} has a median \final{sample} length of only 644 tokens but includes \final{samples} approaching one million tokens. Because dense attention FLOPs grows quadratically with sequence length, this skew translates into much larger variation in attention.

\begin{figure}[!t]
    \centering
    \includegraphics[width=0.9\linewidth]{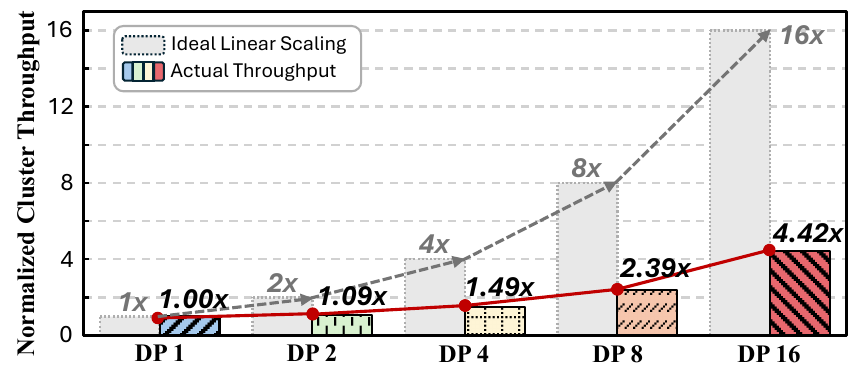}
    \caption{Training throughput of \final{Qwen3-30B-A3B} with 1M-token packed sequences as DP scales from 1 to 16.}
    \label{fig:intro}
    \vspace{-10pt}
\end{figure}

Sequence packing \compresses{preserves} this variation. Packing multiple samples into fixed-token sequences balances activation memory and \compresses{linear-cost} operators, but the attention cost of a \final{packed sequence} is proportional to the sum of its constituent \final{samples}' squared lengths. \compresses{With} the rest of a \final{packed sequence} fixed, one 40K-token \final{sample} contributes roughly $10\times$ the attention FLOPs of ten 4K-token \final{samples} despite occupying \compresses{the same token count}. Equal token counts can therefore conceal substantially different attention workloads.

This imbalance becomes a cluster-wide performance problem. A slow data-parallel (DP) replica delays the entire synchronization group, while latency variation across microbatches enlarges pipeline-parallel (PP) bubbles. In \final{the Qwen3-30B-A3B} workload with 1M-token packed sequences and context parallel degree \textcolor{black}{CP$=16$}, increasing DP from 1 to 16 and GPUs from 16 to 256 improves cluster throughput by only $4.42\times$, far from the ideal (Figure~\ref{fig:intro}). The resulting scaling efficiency is only $27.6\%$: attention skew at a few workers leaves resources idle throughout the job.

Existing approaches to cross-CP-group imbalance largely operate at two extremes. 
\textit{1)} Data schedulers and adaptive parallelization schemes balance \compresses{\final{samples}, \final{packed sequences}, or microbatches}~\cite{longalign2024,wlbllm2025,ChunkFlow2025,skrull2025,flexsp2025,dcp2025,dynamiccp2026,zhang2026helixpipe}. At these granularities, an outlier can still dominate \compresses{one CP group's workload}; compensating for it through variable-length packing can also increase the token and activation-memory budgets. 
\textit{2)} At the other extreme, DistCA~\cite{DistCA2025} disaggregates attention over a global worker pool. A global pool maximizes task-placement scope, but its communication domain grows with the cluster and can span low-bandwidth inter-supernode links. These limitations expose \compresses{our central design question}: \emph{over what scope should attention be balanced?}

Our key insight is to use the law of large numbers (LLN) as a scaling principle for load balancing in long-context LLM training: the attention-balancing pool need not grow with the data-parallel degree. 
We define a \emph{sequence pool} as a fixed-size group of packed sequences, together with the CP groups processing them, over which attention work may be exchanged (i.e., \emph{attention pool}). 
Aggregating multiple sequences causes a pool's average attention workload to concentrate around the dataset mean. 
Consequently, the required pool size is primarily determined by the workload distribution and the target imbalance, rather than scaling proportionally with the DP degree.
As DP scales out, \sysname increases the \emph{number} of fixed-size sequence pools instead of the \emph{size} of each pool. It bounds the scope of every attention exchange, naturally supports DP weak scaling, and enables the cluster scheduler to place each pool within a concentrated domain.

\final{However, bare LLN alone} is insufficient at the moderate pool sizes required for communication efficiency. The standard pool construction can still leave substantial residual skew. 
\sysname introduces \emph{Variance-Reduced Sequence Placement} (VRSP), which assigns \final{packed sequences} with complementary attention FLOPs to \final{the} same sequence pools. VRSP makes LLN-guided pooling effective at practical pool sizes.
Even with balanced sequence pools, GPUs within each pool can remain imbalanced. \sysname therefore introduces \emph{Tiled Attention Pooling} (TAP), which decomposes attention along both the sequence and head dimensions. Each resulting SH-Tile covers a sequence range and a head range and becomes an independently schedulable attention task. 
\rev{The head dimension increases the scheduling space without forcing smaller sequence blocks, allowing TAP to reduce FlashAttention performance variation while enlarging the placement space. }
A FLOPs-aware placer assigns SH-Tiles across the pool to equalize per-GPU computation. Because tile exchange introduces QKV-dispatch and output-return communication, the \emph{TAP Pipeliner} further divides the exchange into chunks and overlaps it with FlashAttention computation.

\sysname exposes a drop-in context-parallel attention operator and a pluggable data sampler, requiring no changes to model layers, optimizers, gradient accumulation, checkpointing, or pipeline schedules. An asynchronous CPU planner prepares upcoming tile-placement plans while GPU training proceeds. \wyRe{On three production Qwen3 models---a dense 8B and two MoE models (30B-A3B and 235B-A22B)---trained on 256K- and 1M-token production workloads, \sysname improves end-to-end training throughput over WLB-LLM by $44\%$ on average and up to $68\%$ at 256 GPUs, with the margin widening as training scales out.}
\sysname has also been deployed in \final{large-scale training jobs} spanning 32K to 1M tokens and thousands of GPUs, accumulating hundreds of thousands of GPU-hours while preserving training semantics.
This paper makes the following contributions:

\begin{itemize}
\setlength{\itemsep}{2pt}
\setlength{\topsep}{2pt}
\setlength{\parsep}{0pt}
\setlength{\partopsep}{0pt}
\item We introduce LLN-guided sequence pooling, a scaling principle in which scaling-out DP increases the number of fixed-size sequence pools rather than the size of each pool, bounding the attention-exchange domain.
\item We design VRSP to make LLN-guided pooling effective for finite, long-tailed workloads by explicitly reducing inter-pool FLOPs skew.
\item We design TAP with sequence~$\times$~head tiling for intra-pool balance, together with a pipelined runtime that overlaps tile exchange and attention computation.
\end{itemize}
\vspace{-3pt}

%% file: 2-background.tex
\section{Background}
\label{sec:Background}

\subsection{Parallelism for Long-Context LLM Training}
\label{sec:bg:parallelism}

\textbf{Data Parallel (DP).}
\compresses{A \emph{DP replica} denotes} a logical execution lane on the attention path: each replica processes a disjoint portion of the global batch, and replicas synchronize gradients at optimizer boundaries via AllReduce or ReduceScatter~\cite{zero2020, fsdp2023}.
The slowest rank stalls the \compresses{whole} group \compresses{per} \restruct{synchronization}.
\textbf{Pipeline Parallel (PP).}
PP partitions model layers into sequential stages~\cite{huang2019gpipe,harlap2018pipedream,li2025slimpipe,liu2025mario,wang2026adahc} fed by microbatches; unequal per-microbatch workloads introduce pipeline bubbles that idle the remaining stages.
\restruct{In real deployments, DP replicas and PP stages may span supernodes.}

\textbf{Context Parallel (CP).}
CP distributes the computation and activation state of a single packed sequence across a group of workers, enabling long sequences that would be impractical to process on one GPU~\cite{korthikanti2023reducing}.
Existing schemes partition attention along the sequence axis (e.g., Ring Attention~\cite{ring_attention2023}) or the head axis (e.g., Ulysses~\cite{ulysses2023}).
\restruct{As these schemes target core attention, we use CP hereafter as a paper-specific abstraction for \compresses{workers that jointly execute} \compresses{core} attention of one packed sequence; \compresses{it} also subsumes the tensor-parallel degree \compresses{on} the attention path.
As CP collectives lie on the critical path of each attention call, CP groups are typically mapped to high-bandwidth interconnect domains; across groups, however, per-sequence FLOPs differ by orders of magnitude, leaving a per-worker attention-FLOPs imbalance that motivates this paper (quantified in \S\ref{sec:motivation:imbalance}).}

\textbf{Notation.}
We denote the parallelism dimensions as $\text{DP}$, $\text{PP}$, and $\text{CP}$, with total workers {$N=\text{DP}\times\text{PP}\times\text{CP}$}, and global batch size and microbatch size as $\text{GBS}$ and $\text{mbs}$, respectively\footnote{Tensor parallelism on the attention path is included in CP. In \final{typical} MoE deployments, expert parallelism is overlaid on the $\text{DP}\times\text{CP}$ device mesh~\cite{gale2023megablocks,zhai2023smartmoe} rather than introduced as an additional multiplicative dimension. Accordingly, a DP replica denotes a logical attention-path replica and does not imply that every expert parameter is fully replicated.}.
\wy{We set $\text{mbs}{=}1$ throughout: a single packed long sequence already strains per-device memory and supplies ample compute intensity per microbatch, so batching more sequences is neither affordable nor necessary.}
Thus, $\text{GBS}$, $\text{mbs}$, and the gradient-accumulation factor $\text{GA}$ count packed sequences, and each DP replica processes $\text{GA}=\text{GBS}/\text{DP}$ packed sequences between optimizer updates.
For a fixed gradient-accumulation index $i$, the packed sequences assigned to that index across all DP replicas form a \emph{microbatch group}.

\subsection{Attention Mechanism}
\label{sec:bg:attn}

\wy{This paper focuses on dense causal self-attention within each sample, leaving sparse or linear attention variants to future work.
Multi-head self-attention (MHA)~\cite{attentionisallyouneed2017} consists of QKV projection, core attention, and output projection.
QKV and output projections are linear layers with $\mathcal{O}(l)$ cost in sequence length $l$ and are negligible in the long-context regime; core attention computes pairwise token interactions and dominates with $\mathcal{O}(l^2)$ cost.} For one head,
\begingroup
\setlength{\abovedisplayskip}{4pt}
\setlength{\belowdisplayskip}{4pt}
\begin{equation}
\small
\texttt{CoreAttn}(\mathbf{Q},\mathbf{K},\mathbf{V};\mathbf{M})
= \mathrm{softmax}\!\left(\mathbf{QK}^T / \sqrt{d_h} + \mathbf{M}\right)\mathbf{V}.
\end{equation}
\endgroup
where $\mathbf{Q}, \mathbf{K}, \mathbf{V} \in \mathbb{R}^{l \times d_h}$ are the query, key, and value matrices, and $\mathbf{M} \in \mathbb{R}^{l \times l}$ is the causal attention mask.

\restruct{Core attention is parameter-free: its output depends only on the Q/K/V tensors, mask, and execution metadata, so any compatible worker can execute it, enabling the cross-worker workload migration used by \sysname (\S\ref{sec:design:tap}).}

\subsection{Sequence Packing}
\label{sec:bg:packing}

Sequence packing~\cite{krell2021packing, packing2025} concatenates multiple training \emph{samples} into a \emph{packed sequence} and applies a block-diagonal causal mask to \compresses{prevent cross-sample information flow}.

\textbf{Fixed-token packing} constructs every packed sequence with the same total token count $L$.
This ensures uniform activation memory and balanced non-attention workloads (linear layers, normalization) across DP replicas.
However, attention FLOPs remain imbalanced: core attention cost is quadratic in per-sample \restruct{length}.
Formally, if a packed sequence $S$ contains samples with lengths $\{\ell_j\}$, we use
\begingroup
\setlength{\abovedisplayskip}{4pt}
\setlength{\belowdisplayskip}{4pt}
\begin{equation}
\small
w(S)\;\propto\;\sum_j \ell_j^2
\end{equation}
\endgroup
as its attention-workload proxy, absorbing the causal factor, number of heads, and head dimension into the proportionality constant. Equal $L$ therefore does not imply equal $w(S)$.

\textbf{Workload-aware variable-token packing}~\cite{wlbllm2025} relaxes the equal-token constraint and allows packed sequences \compresses{different} token counts so as to equalize attention workloads across sequences.
This trades memory balance for FLOPs balance, since variable-length sequences consume different activation memory; more importantly, a single outlier sample may \compresses{exceed} the target workload for \compresses{a} packed sequence, so no arrangement of \compresses{remaining} samples can remove this lower bound.
WLB-LLM~\cite{wlbllm2025} mitigates this by delaying outlier samples to later batches, but \compresses{this} changes the sample multiset contributing to an optimizer \compresses{update, violating} the step-equivalence requirement adopted in this paper.

This paper therefore defaults to fixed-token packing and addresses the residual attention FLOPs imbalance through attention-workload scheduling in \S\ref{sec:design}, an approach orthogonal to data repacking and composable with it.

%% file: 3-motivation.tex
\begin{figure}[!t]
    \centering
    \includegraphics[width=\linewidth]{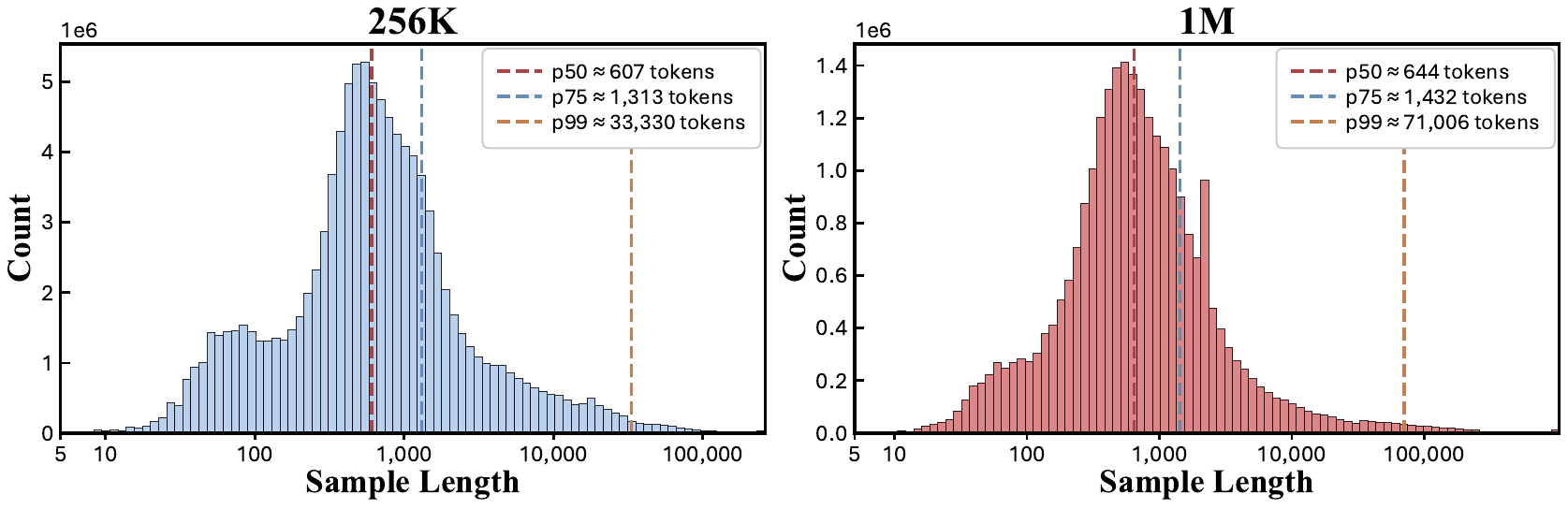}
    \caption{Sample-length distributions before packing on the 256K and 1M production datasets.}
    \label{fig:docmument_length}
    \vspace{-10pt}
\end{figure}

\section{Motivation}
\label{sec:motivation}

\subsection{Attention FLOPs Imbalance under Long-Tailed Distributions}
\label{sec:motivation:imbalance}

Long-context training datasets exhibit strongly long-tailed sample-length distributions: most raw samples \restruct{are} short, while the \final{longest} \restruct{\final{approach}} \final{one million} tokens (\final{Figure}~\ref{fig:docmument_length}).
Under fixed-token packing, however, equal token counts do not imply equal attention workload.  For a packed sequence $S$ \restruct{of} samples with lengths $\{\ell_j\}$, the workload proxy from \S\ref{sec:bg:packing} is $w(S)\propto\sum_j\ell_j^2$.  The sample-length tail therefore induces substantial attention-workload skew among packed sequences and, in turn, among workers.

\textbf{Cross-group imbalance.}
Existing intra-CP dispatchers can handle imbalance within a CP group (\S\ref{sec:bg:parallelism}), but \compresses{do} not remove workload skew across CP groups.  \restruct{Along DP, lighter replicas wait for stragglers at synchronization points: on the 1M workload with CP$=16$, \compresses{scaling} DP from 1 to \final{16} yields only a \final{$4.42\times$} throughput gain (\final{Figure}~\ref{fig:intro}).}  This measurement alone does not attribute the entire scaling gap to attention; \restruct{\sysname's recovery of most lost DP scaling in \wyRe{\S\ref{sec:eval:scaling}} provides} \compresses{complementary indirect} \restruct{evidence}.
Along PP, \compresses{microbatches} in \compresses{one} gradient-accumulation window can carry different attention workloads.  \restruct{Their unequal compute times delay dependent pipeline stages and create compute-induced bubbles even under an unchanged pipeline schedule (\wyRe{\S\ref{sec:eval:scaling}}).}

Our objective is therefore to reduce cross-CP-group, per-worker attention skew and the resulting DP waiting and PP bubbles, rather than force every GPU in every microbatch to match the global mean exactly.

\subsection{Bounded Attention-Workload Redistribution via the Law of Large Numbers}
\label{sec:motivation:scope}

\restruct{Core attention is parameter-free (\S\ref{sec:bg:attn}):} a compatible worker can execute a task given its Q/K/V tensors, mask, and execution metadata.  This property permits \compresses{redistributing attention workloads} across CP groups, but leaves a central question: \restruct{over what scope?}  \restruct{We} define a \emph{sequence pool} at a fixed PP stage and gradient-accumulation (GA) index.  It contains \restruct{the} \compresses{packed} sequences and \restruct{CP} groups of $P$ DP replicas\restruct{, executed} by $P\cdot\mathrm{CP}$ workers, where $P\mid\mathrm{DP}$.  A pool is a spatial redistribution domain: \restruct{GA indices may reuse the same workers, but their runtime tasks form distinct pools.}

\begin{figure}[th]
    \centering
    \includegraphics[width=\columnwidth]{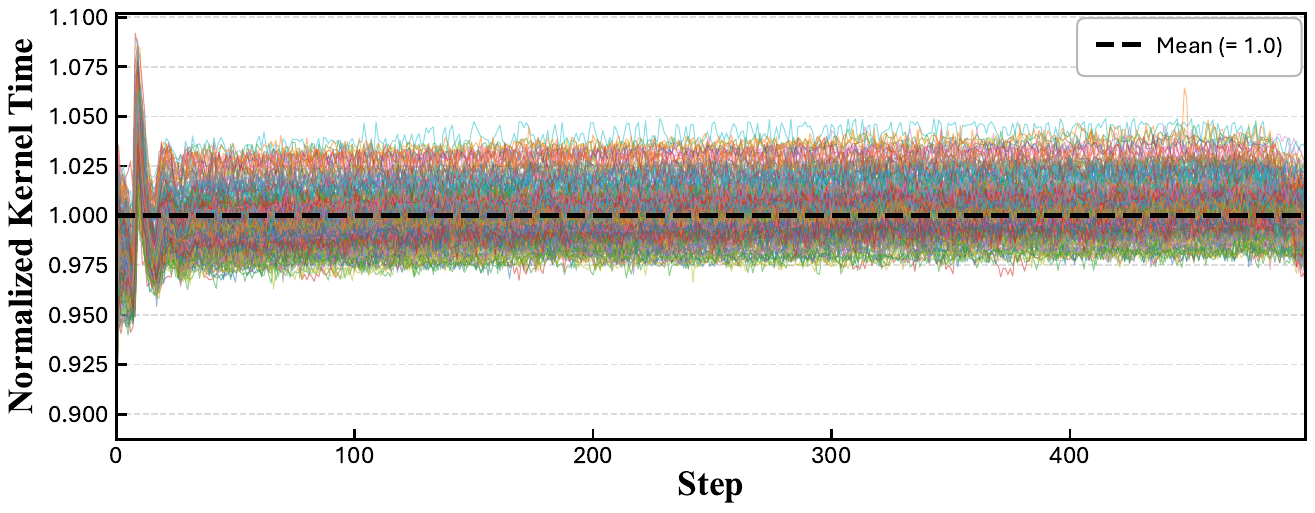}
    \caption{Runtime variation across 256 identical GPUs executing the same 32K causal-attention input for 500 steps.}
    \label{fig:fa_kernel_eff}
\end{figure}

\textbf{Local scope ($P{=}1$).}
\restruct{Without cross-group redistribution, balance \compresses{demands} data placement, repacking, or adaptive parallelism (e.g., WLB-LLM~\cite{wlbllm2025}, FlexSP~\cite{flexsp2025}).}  \compresses{Such} approaches face an indivisible-outlier lower bound: rearranging other samples cannot make a single outlier cheaper.  Methods that defer an outlier to a later optimizer step also change \compresses{its} sample multiset, violating the step-equivalence \restruct{requirement}.

\textbf{Large-scope extreme ($P{=}\mathrm{DP}$).}
\restruct{\compresses{In} the sequence-pool abstraction, the largest scope per PP stage spans all DP replicas; DistCA}~\cite{DistCA2025} \restruct{extends \compresses{past} this in-model extreme, sending} core-attention tasks from DP/PP workers to a dedicated attention-server pool.  This \compresses{scope} \compresses{aids} workload aggregation but \compresses{has} three costs.  \restruct{First}, its attention communication domain grows with the cluster and can cross low-bandwidth inter-supernode links.  Second, a larger pool averages more data-workload variation but \restruct{sees more device-runtime variation}, so \compresses{a FLOPs-balanced assignment need not be time-balanced}.  Third, DistCA-style cross-PP disaggregation \compresses{needs} dedicated ping-pong execution and runtime co-design; it is \compresses{no} pipeline-schedule-transparent local replacement.

Figure~\ref{fig:fa_kernel_eff} \compresses{shows} the second cost.  Under \compresses{this} protocol, the per-step spread averages 7.01\%\compresses{, reaching} 15.07\%.  \restruct{Thus}, enlarging a pool trades better workload aggregation for a wider communication domain and greater runtime-variability exposure, suggesting a finite effective \restruct{scope}.

\textbf{Concentration in fixed-size pools.}
\restruct{Let $X=w(S)$ denote packed-sequence workload with mean $\mu$ and standard deviation $\sigma$ ($\mathrm{CV}=\sigma/\mu$); within a GBS window, $D=\mathrm{GBS}$ sequences form $K=D/P$ pools of $P$ sequences each, with pool sums $L_k$ and residual imbalance $R-1$ for $R=\max_k L_k/(P\mu)$.  Under the analytical approximation of random, independent and identically distributed (IID) grouping with finite variance, the Central Limit Theorem (CLT) makes $L_k$ approximately Gaussian; normalizing by $P\mu$ gives standard deviation $\mathrm{CV}/\sqrt{P}$, and approximating the expected maximum of the $K$ normalized sums yields}
\begin{equation}
\small
\mathbb{E}[R] \;\approx\;
1+\frac{\mathrm{CV}}{\sqrt{P}}\sqrt{2\ln K}
=
1+\frac{\mathrm{CV}}{\sqrt{P}}\sqrt{2\ln(D/P)}.
\label{eq:lln-theory}
\end{equation}

\begin{figure}[!t]
    \centering
    \includegraphics[width=\columnwidth]{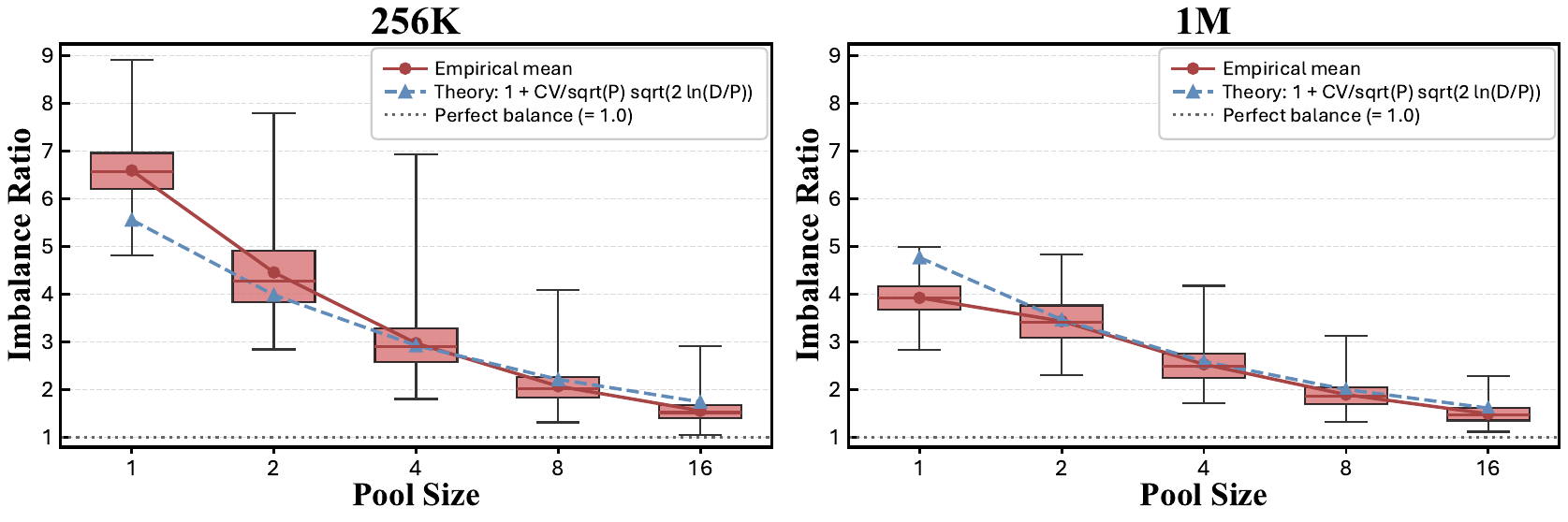}
    \caption{\wy{Residual inter-pool imbalance versus pool size on the 256K (left) and 1M (right) datasets (GBS$=128$).  Distributions from empirical \rev{production-order} grouping; markers show means; the analytical curve follows Equation~\ref{eq:lln-theory}.}}
    \label{fig:imbalance_at_diff_pool_size}
\end{figure}

\restruct{Figure}~\ref{fig:imbalance_at_diff_pool_size} evaluates practical sizes $P\le16$\restruct{: at} $P=8$, \rev{production-order} grouping still leaves $R-1\approx1.08$ on 256K and $R-1\approx0.90$ on 1M.  \restruct{The LLN explains why \compresses{a} fixed-size pool's concentration is set by $P$ and the workload distribution rather than \compresses{$K$}, yet the cluster-wide maximum retains the $\sqrt{\ln K}$ term, so bare LLN converges too slowly at practical $P$.}

\textbf{Scaling principle: fixed-size, DP-only pools.}
\restruct{The attention-balancing pool therefore need not grow with the DP degree: pools remain fixed in size, and DP scale-out increases their number rather than their size, bounding every redistribution domain.  Suppressing the $\sqrt{\ln K}$ residual across pools requires finite-window inter-pool placement, while intra-pool worker balancing addresses skew within each pool.}

\subsection{Head-Axis Splitting for Communication Overlap}
\label{sec:motivation:head}

\begin{figure}[htbp]
    \centering
    \includegraphics[width=\columnwidth]{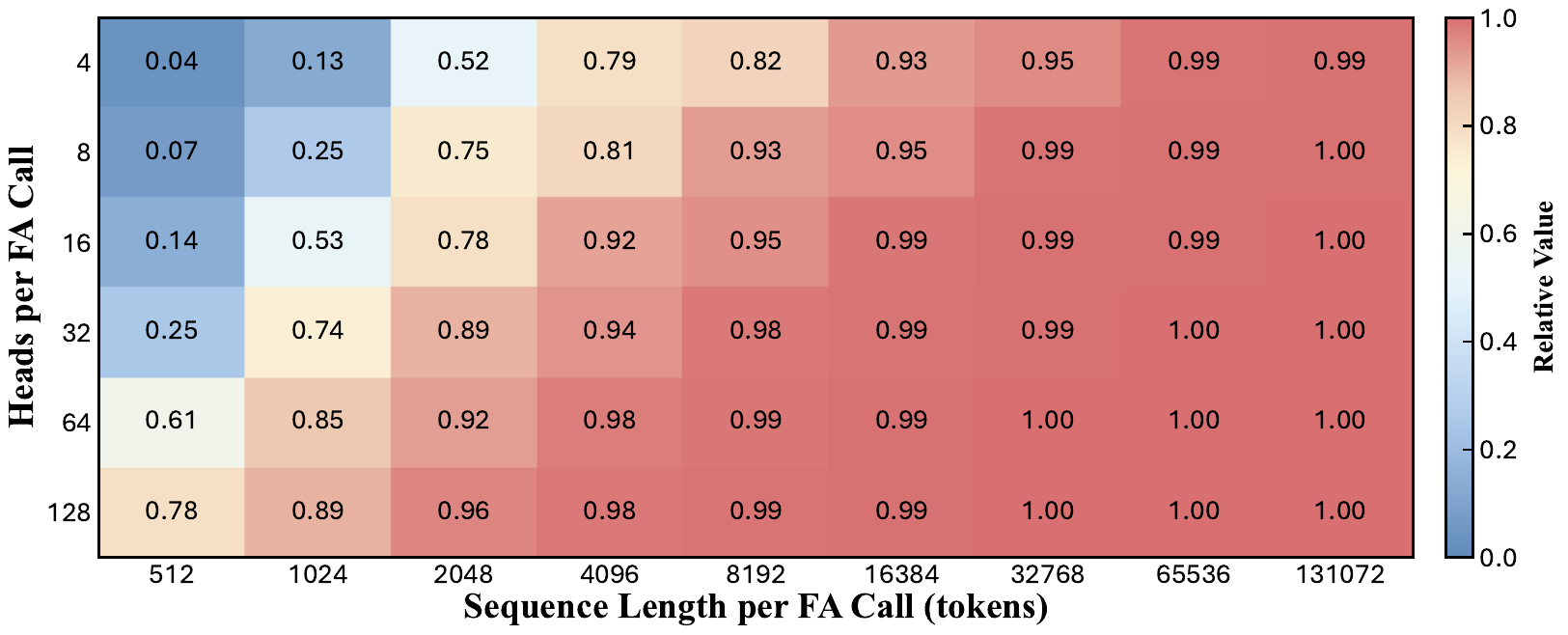}
    \caption{\wy{Measured FlashAttention throughput versus per-call sequence-block length (horizontal axis) and head count (vertical axis), normalized to \restruct{this configuration's peak}.}}
    \label{fig:fa_heatmap}
    \vspace{-10pt}
\end{figure}

Attention-workload redistribution transfers Q/K/V tensors and attention outputs within a pool, introducing communication.  To avoid eroding the gains from load balancing, this communication must be overlapped with attention computation.  Such overlap is most effective when chunks are balanced in both compute and communication.

For the standard dense causal attention considered here, equal head-axis chunks \compresses{share} query--key interactions, Q/K/V/output byte volume, and execution structure.  The conventional sequence-axis alternatives \restruct{\compresses{lack}} both properties: fixed-token chunks become \compresses{costlier} later in the causal order, whereas fixed-FLOPs chunks contain different \compresses{counts} of query tokens and \compresses{hence} different communication volumes\restruct{; \compresses{thus}, of the axes considered here, only the head axis provides} uniform compute and communication chunks.

\restruct{Head-axis splitting also preserves kernel efficiency better in our measured configuration (\final{Figure}~\ref{fig:fa_heatmap}): normalized FlashAttention throughput is less sensitive to reducing heads per call than to shortening sequence blocks.}

%% file: 4-design.tex
\begin{figure*}[t]
  \centering
  \includegraphics[width=\textwidth]{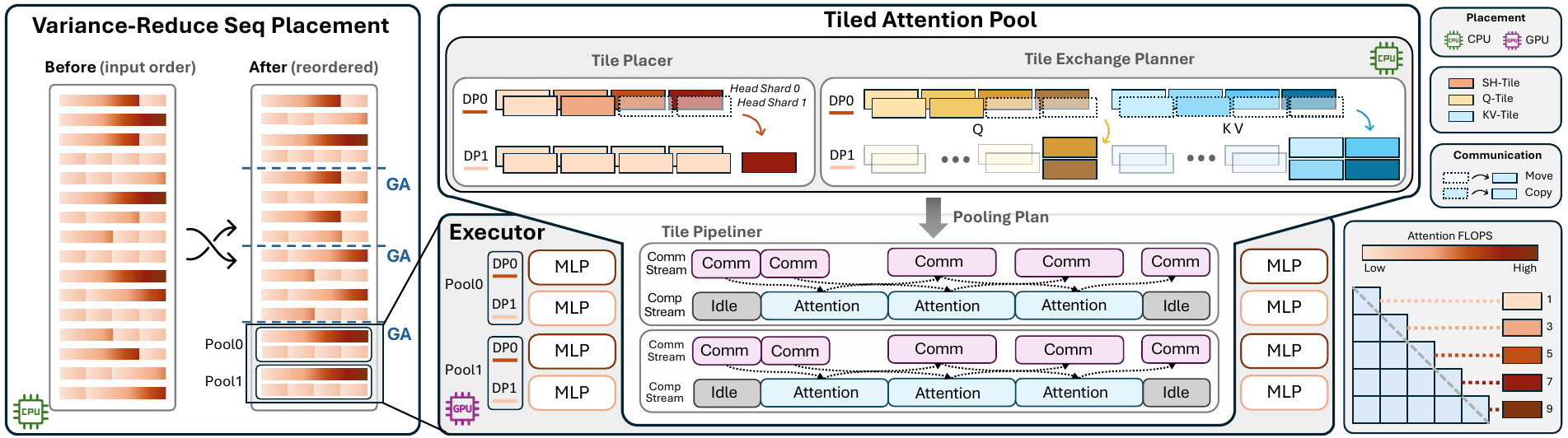}
  \caption{Design overview of \sysname.}
  \label{fig:overview}
\end{figure*}

\section{Design}
\label{sec:design}

\subsection{Overview}
\label{sec:design:overview}

\sysname adopts a two-level load-balancing design.  Its central abstraction is the \emph{sequence pool}: at a fixed PP stage and gradient-accumulation (GA) index, a pool contains the packed sequences of $P$ DP replicas and their CP groups, comprising $W=P\cdot\mathrm{CP}$ workers.  We require $P\mid\mathrm{DP}$ and keep pool membership fixed during training.  Pools span only DP; different GA indices reuse the same worker groups over time but form distinct logical pools.

\sysname uses three components to address two levels of skew.  \emph{Variance-Reduced Sequence Placement} (VRSP; \S\ref{sec:design:vrsp}) balances aggregate attention FLOPs across sequence pools.  \emph{Tiled Attention Pooling} (TAP; \S\ref{sec:design:tap}) \restruct{decomposes each packed sequence into \emph{SH-Tiles}---independently placeable core-attention tasks \compresses{pairing} one sequence block with one head shard---and assigns them} across the $W$ workers in a pool to balance estimated FLOPs.  The \emph{TAP Pipeliner} (\S\ref{sec:design:pipeliner}) \compresses{also} overlaps intra-worker-group tensor transfers with attention \restruct{\compresses{compute}.}

\begin{figure}[!t]
  \centering
  \includegraphics[width=\linewidth]{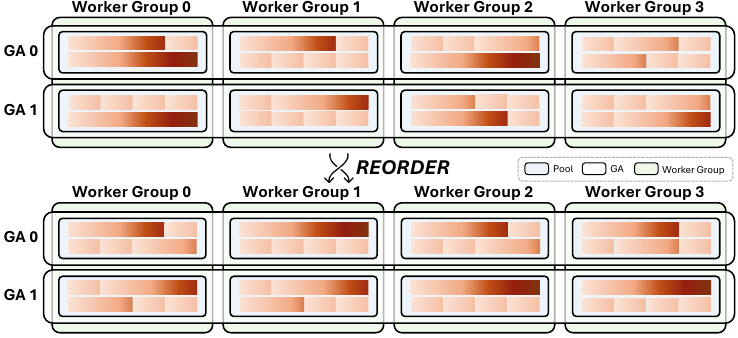}
  \caption{\wy{GA-index/pool layout ($\text{GBS}{=}16$, $\text{DP}{=}8$, $P{=}2$): four pools per GA index span disjoint pairs of DP replicas; the next GA index reuses the same four worker groups.}}
  \label{fig:GA and pool}
  \vspace{-10pt}
\end{figure}

Figure~\ref{fig:overview} shows the workflow.  VRSP runs in the data path and emits the reordered packed sequences.  A CPU planner derives a per-tile assignment and its communication schedule from sequence-length metadata.  The assignment is reused across Transformer layers with the same layout, while each layer transfers its own Q/K/V and output tensors.  \compresses{Thus}, no placement decision is made on the GPU critical path.
Figure~\ref{fig:GA and pool} \compresses{shows the pool layout}.  With $\text{GBS}{=}16$, $\text{DP}{=}8$, and $\text{mbs}{=}1$, the GA window has two indices.  At each index, $\mathrm{DP}/P=4$ pools execute concurrently; across the optimizer-step window, VRSP constructs $\mathrm{GBS}/P=8$ logical \restruct{pools.}

\subsection{Variance-Reduced Sequence Placement}
\label{sec:design:vrsp}

VRSP balances the aggregate workload of all sequence-pool instances in one optimizer-step window.  Its scheduling unit is a complete packed sequence: VRSP neither splits a packed sequence nor changes its internal packing.  For packed sequence $S_i$ with constituent raw-sample lengths $\{\ell_{ij}\}$, it uses
\begin{equation}
F_i=w(S_i)\propto\sum_j\ell_{ij}^{2}.
\label{eq:vrsp-work}
\end{equation}
Given $\mathrm{GBS}$ packed sequences and pool size $P$, VRSP forms $K=\mathrm{GBS}/P$ sequence-pool instances, each containing exactly $P$ packed sequences.  It may reorder packed sequences across GA indices, DP replicas, and pools, but preserves the raw-sample multiset of the entire optimizer step.

\textbf{Natural concentration is insufficient at bounded pool sizes.}
The LLN-guided analysis in \S\ref{sec:motivation:scope} explains why increasing $P$ concentrates a randomly constructed pool's aggregate workload.  Yet production training presents a finite GBS window, and its packed-sequence workloads remain long-tailed.  Consequently, \rev{production-order} grouping leaves a large cluster-wide maximum at the moderate $P$ required to bound communication \restruct{(\S\ref{sec:motivation:scope}); even at $P{=}64$, $R{-}1$ remains $0.09$ (256K) and $0.08$ (1M)}.  Relying on natural concentration alone would therefore require nearly collapsing the GBS window into one communication pool.

\textbf{Heavy--light grouping balances the realized pool workloads.}
VRSP uses the placement freedom within one optimizer step instead of enlarging the communication domain\restruct{: because} VRSP cannot reduce an outlier's cost, it places heavy packed sequences \compresses{with} lighter ones so that each pool's aggregate workload approaches the mean across \compresses{the} $K$ \restruct{pools}.

\textbf{Exact-cardinality greedy placement.}
VRSP must jointly satisfy two requirements: balance aggregate FLOPs and assign exactly $P$ packed sequences to every pool.  As shown in \final{Algorithm}~\ref{alg:vrsp}, VRSP adapts the \emph{longest-processing-time-first} (LPT) heuristic~\cite{graham1969bounds}\restruct{---assigning tasks heaviest-first, each to the currently least-loaded destination---with an exact-cardinality constraint: it sorts packed sequences by decreasing $F_i$ and repeatedly assigns the next sequence to the least-loaded pool that \compresses{has} fewer than $P$ sequences, removing full pools from the candidate heap}.  This heaviest-first order places difficult outliers while all pools remain available and lets later, lighter sequences fill the residual gaps.  Equal-workload sequences are ordered by \compresses{original} indices; equal-load pools are ordered by pool index, making the procedure deterministic.

\begin{algorithm}[h]
  \caption{Variance-Reduced Sequence Placement}
  \label{alg:vrsp}
  \footnotesize
  \begin{algorithmic}[1]
    \STATE {\bfseries Input:} packed sequences $\{S_i\}_{i=0}^{\mathrm{GBS}-1}$; pool size $P$
    \STATE {\bfseries Output:} reordered sequence list $S'$
    \STATE $F_i\leftarrow\sum_j\ell_{ij}^{2}$ for every $S_i$
    \STATE $K\leftarrow\mathrm{GBS}/P$; initialize $K$ empty pools
    \STATE order $\leftarrow$ sort $(F_i,i)$ by decreasing $F_i$, then increasing $i$
    \STATE $\mathcal H\leftarrow$ min-heap of $(\mathrm{load}_k,k)$ over non-full pools
    \FOR{$i$ {\bfseries in} order}
      \STATE $(\mathrm{load}_k,k)\leftarrow\operatorname{pop}(\mathcal H)$
      \STATE append $S_i$ to pool $k$; $\mathrm{load}_k\mathrel{+}=F_i$
      \IF{pool $k$ contains fewer than $P$ sequences}
        \STATE push $(\mathrm{load}_k,k)$ into $\mathcal H$
      \ENDIF
    \ENDFOR
    \STATE $S'\leftarrow$ concatenate pools in GA-major, pool-minor order
    \STATE {\bfseries return} $S'$
  \end{algorithmic}
\end{algorithm}

\begin{figure}[t]
  \centering
  \includegraphics[width=\columnwidth]{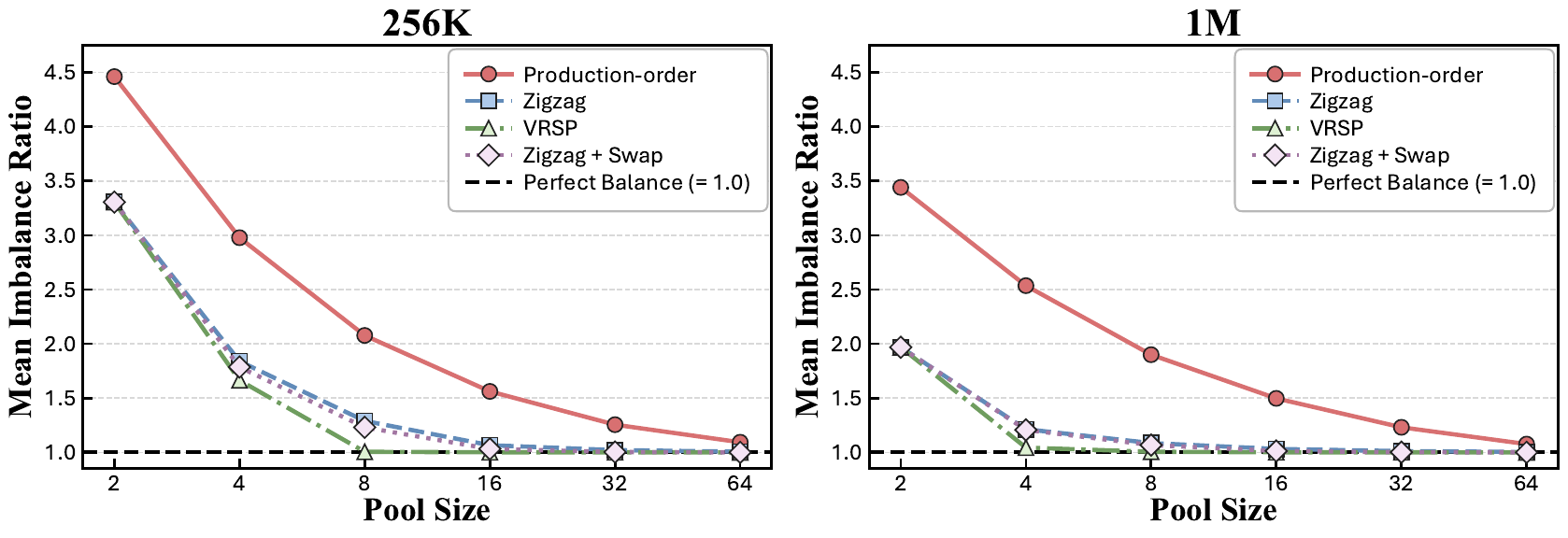}
  \caption{Inter-pool imbalance versus pool size $P$ on the \final{256K and 1M} datasets ($\mathrm{GBS}{=}128$).}
  \label{fig:methods_comparison}
\end{figure}

Writing $D=\mathrm{GBS}$, the procedure sorts the $D$ packed sequences and performs one heap operation per sequence, taking $O(D\log D+D\log K)=O(D\log D)$ time and $O(K)$ pool-state space\rev{---measured at $48$\,$\mu$s and $106$\,$\mu$s per window for $D{=}128$ and $D{=}256$ ($P{=}8$, one CPU core)}.  We use it as a practical exact-cardinality heuristic and do not \restruct{claim} the approximation guarantee of unconstrained classical \restruct{LPT.}
\restruct{Figure~\ref{fig:methods_comparison} shows the reduction: at $P{=}8$, VRSP lowers $R{-}1$ below $0.7\%$ on both datasets, versus $90\%$ and $108\%$ under \rev{production-order} grouping, and outperforms Zigzag and Zigzag-with-swap at all evaluated pool sizes.  VRSP thus supplies the per-step placement mechanism that bare LLN lacks.}

\subsection{Tiled Attention Pooling}
\label{sec:design:tap}

Even after VRSP balances aggregate pool workloads, workers within a pool can remain imbalanced.  TAP decomposes core attention into SH-Tiles, estimates their FLOPs, and assigns them across the pool while accounting for the communication induced by migration.

\subsubsection{SH-Tiles and KV Groups}
\label{sec:design:tap:shtile}

TAP divides each packed sequence at global token positions into blocks of at most $B$ tokens; it does not introduce additional cuts at raw-sample boundaries.  Thus, a block may intersect multiple samples, and the final block may contain fewer than $B$ tokens.  TAP also splits the $h_q$ query heads into $H$ equal-width shards, where $H\mid h_q$.  $H$ is fixed across pools, layers, and packed sequences in one training configuration.  We use the SH-Tile as the smallest independently placeable unit.
\begin{equation}
\mathrm{SH\mbox{-}Tile}\ t=(\texttt{sequence\_block}\ b,\texttt{head\_shard}\ h)
\label{eq:shtile}
\end{equation}

A SH-Tile remains intact even when block $b$ crosses sample boundaries: one variable-length attention call processes its sample fragments under the block-diagonal causal mask.  Let block $b$ cover sample-local query positions $a_{bj},\ldots,e_{bj}-1$ in every intersected raw sample $j$.  TAP estimates the tile's attention FLOPs as Equation~\ref{eq:tile-work}.
\begingroup
\setlength{\abovedisplayskip}{4pt}
\setlength{\belowdisplayskip}{4pt}
\begin{equation}
\small
f_{b,h}\propto
\frac{h_q}{H}
\sum_{j:b\cap j\ne\emptyset}
\sum_{q=a_{bj}}^{e_{bj}-1}(q+1)
\label{eq:tile-work}
\end{equation}
\endgroup
The estimate counts actual causal query--key pairs \restruct{and thus charges no internal padding}.  The placer balances this analytical FLOPs proxy.  Because it depends only on sequence metadata and the fixed tile configuration, the CPU planner can compute it cheaply and deterministically before execution, without online \restruct{profiling}.  Different head shards of the same block have equal $f_{b,h}$ but no KV affinity.

KV reuse instead follows raw-sample and head-shard identity.  A \emph{KV group} is $g=(j,h)$: the complete K/V tensors of raw sample $j$ on head shard $h$.  Multiple query blocks of the same sample and head shard share this KV group.  Even when a query block needs only a causal prefix, the runtime fetches the complete group and applies the causal mask during attention.  A cross-sample SH-Tile therefore references the set in Equation~\ref{eq:tile-kvset}.
\begin{equation}
\mathcal K(t)=\{(j,h)\mid j\text{ intersects tile }t=(b,h)\}
\label{eq:tile-kvset}
\end{equation}
KV groups therefore serve only as reuse information during placement.  The placer still assigns each complete SH-Tile exactly once, to one worker.

\subsubsection{Communication-Aware Tile Placement}
\label{sec:design:tap:placer}

Each SH-Tile $t$ has one Q-home worker $q(t)$, which supplies its Q tensor and receives its output.  The KV groups referenced by $t$ have independent source mappings and may be assembled from fragments held by multiple workers.  The Q-home therefore need not hold the tile's complete K/V.

The placer tracks, for every destination worker $r$, its estimated load $L_r$ and the set $\mathcal C_r$ of KV groups already selected for that destination.  Because a fetched group remains resident until the current layer's pool execution completes, another tile at the same destination reuses it without a second transfer.  Placing tile $t$ at $r$ adds the communication volume as Equation~\ref{eq:placement-comm}.
\begingroup
\setlength{\abovedisplayskip}{4pt}
\setlength{\belowdisplayskip}{4pt}
\begin{equation}
\small
\Delta_{\mathrm{comm}}(t,r)=
\sum_{g\in\mathcal K(t)\setminus\mathcal C_r}V_g+
\mathbf 1[r\ne q(t)]\bigl(V_Q(t)+V_O(t)\bigr)
\label{eq:placement-comm}
\end{equation}
\endgroup
where $V_g$, $V_Q(t)$, and $V_O(t)$ are the actual KV, Q, and output byte volumes.  Selecting the Q-home avoids both Q dispatch and output return; selecting a destination that already holds a referenced KV group avoids that group's KV transfer.

\begin{algorithm}[h]
  \caption{Communication-Aware SH-Tile Placement}
  \label{alg:placer}
  \footnotesize
  \begin{algorithmic}[1]
    \STATE {\bfseries Input:} tiles $\mathcal T$; workers $\mathcal W$; $\tau=0.03$
    \STATE {\bfseries Output:} assignment $\sigma$
    \STATE $C\leftarrow(1+\tau)(\sum_{t\in\mathcal T}f_t)/|\mathcal W|$
    \STATE order $\leftarrow$ sort tiles by decreasing $f_t$, then tile ID
    \FOR{$t$ {\bfseries in} order}
      \STATE $\mathcal F\leftarrow\{r\in\mathcal W:L_r+f_t\le C\}$
      \IF{$\mathcal F\ne\emptyset$}
        \STATE $r^*\leftarrow\arg\min_{r\in\mathcal F}
          (\Delta_{\mathrm{comm}}(t,r),L_r,r)$
      \ELSE
        \STATE $r^*\leftarrow\arg\min_{r\in\mathcal W}
          (L_r,\Delta_{\mathrm{comm}}(t,r),r)$
      \ENDIF
      \STATE $\sigma(t)\leftarrow r^*$; $L_{r^*}\mathrel{+}=f_t$
      \STATE $\mathcal C_{r^*}\leftarrow\mathcal C_{r^*}\cup\mathcal K(t)$
    \ENDFOR
    \STATE {\bfseries return} $\sigma$
  \end{algorithmic}
\end{algorithm}

\begin{figure}[!t]
  \centering
  \includegraphics[width=\columnwidth]{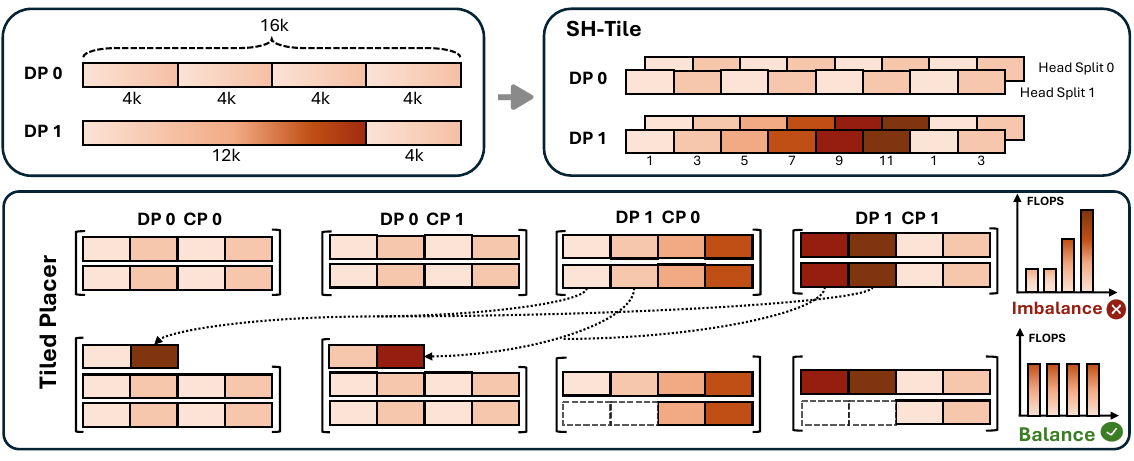}
  \caption{SH-Tile construction and communication-aware placement (take $B{=}2048$, $H{=}2$, $P{=}2$, $\mathrm{CP}{=}2$) for example.}
  \label{fig:tile-placer}
\end{figure}

\restruct{Figure~\ref{fig:tile-placer} illustrates the construction and placement; Algorithm~\ref{alg:placer} formalizes the rule: the placer assigns each tile to a destination that \compresses{stays} below the soft load target $C$, minimizing incremental communication, then \compresses{load}, then worker ID; if no such destination exists, a least-loaded worker accepts the tile, so every input \compresses{gives} a complete assignment.}  $C$ guides the balance--communication trade-off rather than providing a worst-case load guarantee; in particular, a tile can be individually larger than the remaining headroom of every worker.  We fix $\tau=0.03$ in all configurations\restruct{, with tile ordering and destination scoring executed on CPU}.

\subsubsection{Tile Exchange Planning}
\label{sec:design:tap:exchange}

The planner converts $\sigma$ \compresses{to} tensor transfers.  \restruct{KV} transfer is deduplicated \compresses{per} sample, head shard, and destination worker: \compresses{tiles} at one destination \compresses{referencing} the same KV group share one complete fetch, while two destinations fetch separate copies.  \restruct{Local} Q/output paths and \compresses{resident} KV groups \compresses{need} no \restruct{transfer.}
\rev{The backward pass reuses the same assignment with reversed transfers: the output gradient follows the forward Q-dispatch path to each tile's destination, the query gradient returns along the forward output path, and KV gradients \compresses{retrace} the KV-fetch paths.  Backward communication \compresses{thus equals} the forward pass \compresses{in volume} and needs no separate plan.}

\subsection{TAP Pipeliner}
\label{sec:design:pipeliner}

\begin{figure}[!t]
  \centering
  \includegraphics[width=\columnwidth]{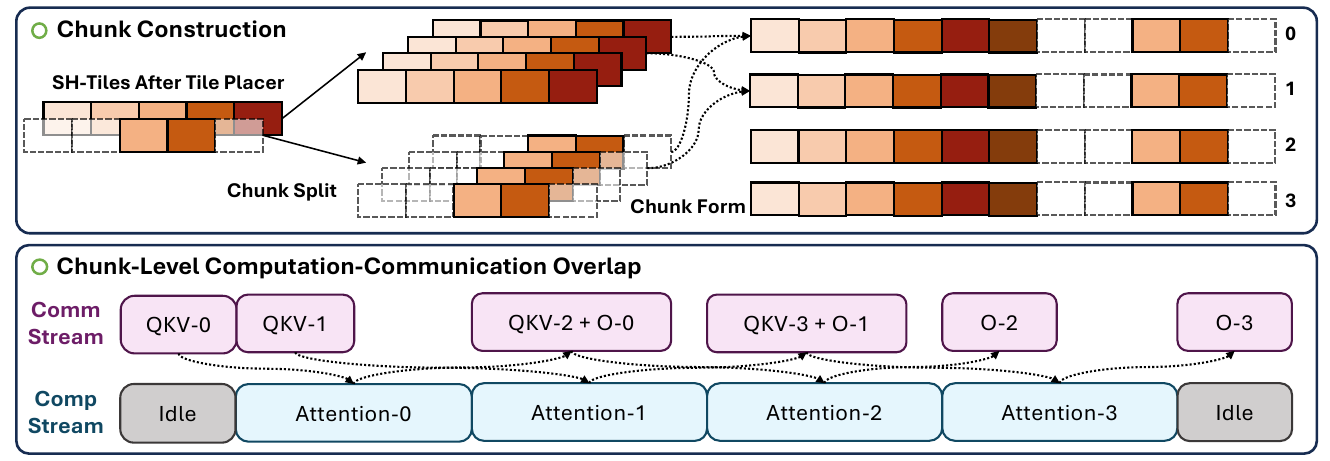}
  \caption{TAP Pipeliner with $M{=}4$.  \compresses{Nonblocking dispatch and return overlap computation} of adjacent equal-head chunks; the first dispatch and final return \compresses{stay} exposed.}
  \label{fig:pipeliner}
  \vspace{-10pt}
\end{figure}

TAP's tensor movement can offset the gain from balancing unless it overlaps attention computation.  The Pipeliner exploits the head axis to create chunks that contain proportional pieces of every tile assigned to a worker \final{(Figure~\ref{fig:pipeliner})}.

\textbf{Equal-head chunk construction.}
Every assigned SH-Tile has width $h_q/H$ heads.  The executor first concatenates the worker's SH-Tiles along the sequence/variable-length batch dimension, producing an aggregate workload whose head width remains $h_q/H$.  It then divides this head dimension into $M$ equal pieces, with $M\mid(h_q/H)$.  \restruct{Chunk $m$ therefore contains the $m$-th head piece of \emph{every} assigned tile; every chunk thus contains} the same fraction of each tile and hence the same estimated FLOPs, Q/output bytes, KV bytes, and execution structure.  The logical KV fetch of each group is correspondingly transferred in $M$ head pieces.  We use the configurable default $M=4$ in all experiments.

\textbf{Asynchronous overlap.}
\restruct{The runtime issues all transfers nonblocking.}  \restruct{While} computing an arrived chunk~$i$, it initiates chunk~$i{+}1$'s Q/K/V dispatch and chunk~$i{-}1$'s output return.  It waits for an input handle only before computing the corresponding chunk and waits for outstanding return handles only before assembling the final output.  \restruct{Thus only} chunk~0's dispatch and chunk~$M{-}1$'s return \restruct{are exposed; intermediate transfers overlap compute.}  When each chunk's attention time covers the concurrent transfer time, up to $(M-1)/M$ of communication can be hidden.  

%% file: 5-implementation.tex
\section{Implementation}
\label{sec:impl}

\restruct{\sysname is a 9k-line Python/PyTorch package on an unmodified FlashAttention variable-length kernel.  An offline simulator replays VRSP and TAP \compresses{on} corpus sequence-length metadata to select \compresses{fixed} $(P,H,B)$ (\S\ref{sec:evaluation}).  Each optimizer step, a global sampler runs VRSP, and a per-pool coordinator builds and broadcasts the TAP plan.  The runtime swaps the CP core-attention call for \texttt{\sysid\_attention}, issuing planned Q/K/V dispatches and output returns asynchronously.}

\restruct{Planning runs asynchronously on CPU, hidden behind multi-second GPU steps: workload estimation and VRSP cost $2.2$\,ms (256K, $\mathrm{GBS}{=}128$) and $0.6$\,ms (1M, $\mathrm{GBS}{=}32$) per optimizer step on one core, linear in GBS; TAP placement costs $2.4$\,ms (256K) and $10.7$\,ms (1M) per microbatch per rank; ranks plan in parallel, so DP scale-out adds no per-rank cost.  In production training, VRSP preserves each optimizer step's raw-sample multiset, while TAP runs the same masked core-attention tasks in a potentially different floating-point order; we do not claim bitwise-equivalent trajectories, but deployed runs converge normally.}

%% file: 6-evaluation.tex
\section{Evaluation}
\label{sec:evaluation}

\subsection{Experimental Setup}
\label{sec:eval:setup}

We evaluate \sysname on a production NVIDIA GPU cluster with NVLink intra-node interconnect and RoCE inter-node networking.
\final{All experiments use the latest cluster-compatible FlashAttention~\cite{fa42026} as the core-attention backend and an internal Megatron-LM-style~\cite{shoeybi2019megatron,hu2026tessera} training framework, so performance differences between methods reflect only the mechanisms under evaluation.}

\textbf{Workloads.}
We use two production datasets \compresses{\wyRe{packed to}} sequence lengths of 256K and 1M tokens, whose raw (pre-packing) sample lengths are long-tailed (Figure~\ref{fig:docmument_length}).
We evaluate \sysname in two settings.
(1)~\emph{Microbenchmarks} isolate core attention, capturing \compresses{its} computation and cross-worker communication, \compresses{\wyRe{with}} a synthetic attention head configuration \compresses{and sequence lengths from} these datasets.
(2)~\emph{End-to-end training} trains \wyRe{three production Qwen3 models~\cite{yang2025qwen3}---a dense 8B and two MoE models (30B-A3B and 235B-A22B)} on \wyRe{both} \compresses{datasets}.
\wyRe{Table~\ref{tab:model_config} lists \compresses{all settings' detailed} configurations.}

\begin{table}[t]
\centering
\footnotesize
\caption{\wyRe{Model Configurations.}}
\label{tab:model_config}
\wyRe{\begin{tabular}{@{}lcccccc@{}}
\toprule
Model & $h_q$ & $h_{kv}$ & $d$ & Hidden & FFN & Top-$k$ \\
\midrule
Microbenchmark & 128 & 4 & 256 & -- & -- & -- \\
Qwen3-8B & 32 & 8 & 128 & 4096 & 12288 & -- \\
Qwen3-30B-A3B & 32 & 4 & 128 & 2048 & 768 & 8 \\
Qwen3-235B-A22B & 64 & 4 & 128 & 4096 & 1536 & 8 \\
\bottomrule
\end{tabular}}
\end{table}

\textbf{Comparison methods.}
We compare \sysname with three baselines of increasing balancing \compresses{scope:} \emph{Ulysses}~\cite{ulysses2023}, a widely adopted context-parallelism scheme, ideally balances attention FLOPs within each CP group but leaves \compresses{cross-group imbalance} unaddressed\compresses{;} \emph{WLB-LLM}~\cite{wlbllm2025} \compresses{rebalances across CP groups only through} workload-aware data repacking\compresses{; and} \emph{DistCA}~\cite{DistCA2025} disaggregates core attention onto a cluster-wide attention-server pool \compresses{for global balancing}.
We also ablate \sysname with three cumulative variants: \emph{\sysname (TAP)} enables only Tiled Attention Pooling, \emph{\sysname (TAP+VRSP)} adds Variance-Reduced Sequence Placement, and \emph{\sysname (full)} \compresses{adds} the Pipeliner's communication overlap.
We compare all methods in the isolated core-attention experiments (\S\ref{sec:eval:microbench}), \wyRe{and Ulysses and WLB-LLM end-to-end: applying DistCA to} full training requires invasive \compresses{training-framework modifications}.

\rev{As WLB-LLM has no official release, we adopted the open-source implementation provided by DistCA\footnote{\final{\url{https://github.com/hao-ai-lab/DistCA}}}.}
\restruct{DistCA claims expert-parallelism support, but the released codebase provides none, and its overlap logic is difficult to reconcile with expert-parallel communication, precluding end-to-end evaluation on \final{the} MoE model.
\compresses{Also}, its overlap operates across layers, which the isolated single-layer experiments cannot exercise; we therefore emulate DistCA at SH-Tile granularity, retaining its cluster-wide placement ($P{=}\mathrm{DP}$) and FLOPs-balanced assignment while \compresses{giving} it our Pipeliner overlap.}

\textbf{Metrics.}
\restruct{\sysname targets core-attention FLOPs imbalance; every metric below quantifies the gain from removing it.
The isolated core-attention experiments (\S\ref{sec:eval:microbench}, \S\ref{sec:eval:breakdown}, \S\ref{sec:eval:grid}) measure each worker's complete core-attention time, computation plus communication; the slowest worker sets the step time under bulk-synchronous execution, so we report each step's straggler time by its mean and max across \compresses{one GBS's steps}.}
\wyRe{For end-to-end training (\S\ref{sec:eval:e2e}, \S\ref{sec:eval:scaling}), we report throughput in tokens per GPU per second; for PP, where inter-microbatch imbalance surfaces as waits at DP, EP, and PP synchronization points and cannot be attributed cleanly, we measure each microbatch's forward time within the same pipeline stage.}
Unless otherwise noted, \compresses{results} are normalized to \wyRe{the} Ulysses baseline.

\subsection{\wyRe{End-to-End Performance}}
\label{sec:eval:e2e}

\begin{figure}[t]
    \centering
    \includegraphics[width=\columnwidth]{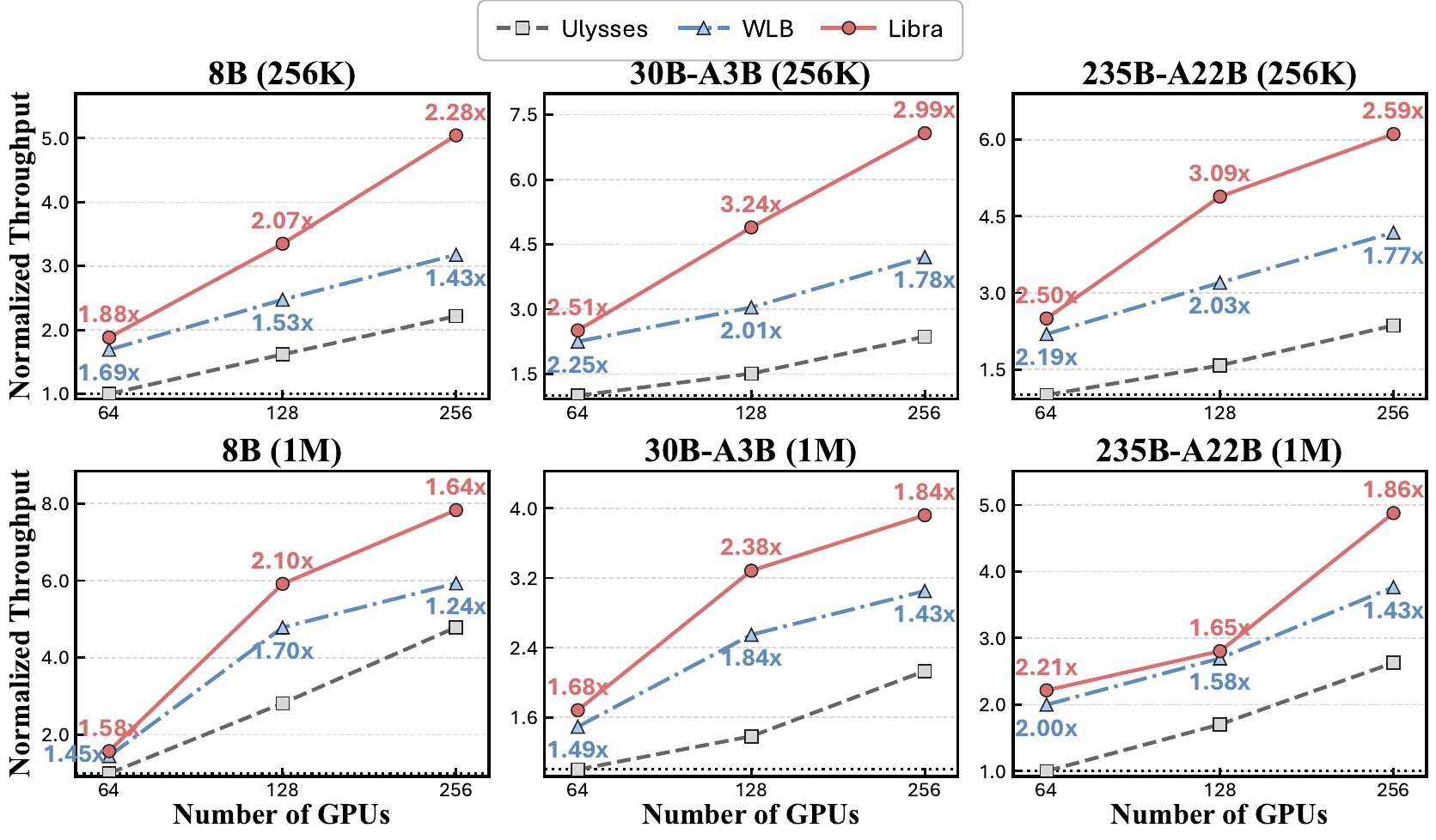}
    \caption{\wyRe{End-to-end training throughput on three production models and both datasets.}}
    \label{fig:e2e_matrix}
\end{figure}

\wyRe{We compare the end-to-end training throughput of \sysname against Ulysses and WLB-LLM on both datasets and the three models in Table~\ref{tab:model_config}, at 64, 128, and 256 GPUs.
On 256K all models run $\text{CP}{=}4$, $\text{PP}{=}4$, $\text{GBS}{=}128$, and on 1M $\text{CP}{=}16$, $\text{PP}{=}2$, $\text{GBS}{=}32$; scaling the GPU count scales only DP ($4{\to}16$ on 256K, $2{\to}8$ on 1M).
Each model packs as many layers as per-GPU memory permits, and the test cluster's limited scale truncates each model to a different extent.
\sysname sets poolsize $P{=}\min(8,\mathrm{DP})$ (256K) and $\min(4,\mathrm{DP})$ (1M) (Figure~\ref{fig:GA and pool}).}

\wyRe{Figure~\ref{fig:e2e_matrix} shows \sysname consistently delivers speedups across all evaluated settings ($1.58\times$--$3.24\times$ over Ulysses at equal GPU count).
At the smallest 64-GPU scale the margin over WLB-LLM is modest ($1.09\times$--$1.14\times$), but as the training scale grows, so do the workload's dynamism and imbalance, and \sysname's advantage amplifies: at 256 GPUs it speeds up training over WLB-LLM by $44\%$ on average and up to $68\%$ (30B-A3B on 256K).
This confirms the advantage \sysname is expected to hold at larger production scales, owing to its data reordering (VRSP) and core-attention pooling (TAP); \S\ref{sec:eval:scaling} further analyzes this scaling advantage.}

\subsection{Attention Balance}
\label{sec:eval:microbench}

\begin{figure}[htbp]
    \centering
    \includegraphics[width=\columnwidth]{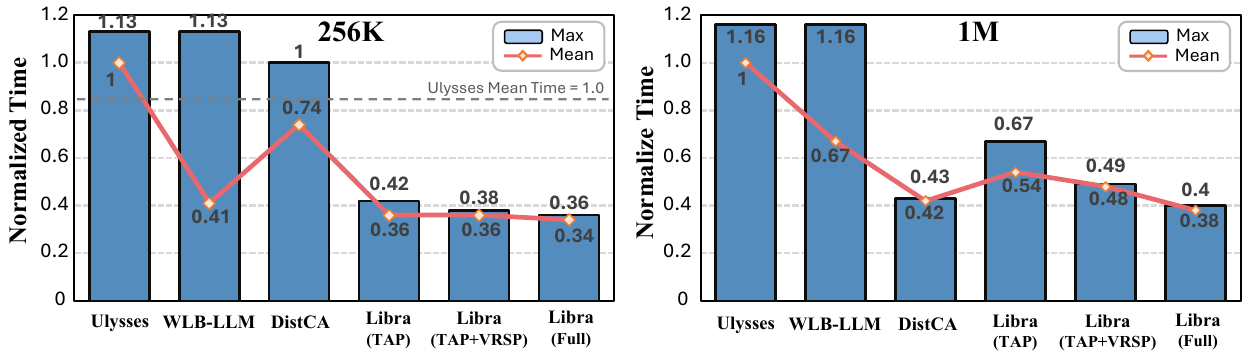}
    \caption{Per-step core-attention latency of \sysname and all baselines on the 256K and 1M datasets.}
    \label{fig:micro_ripple}
\end{figure}

We isolate core attention and compare \sysname against all baselines to directly expose their \compresses{gains}.
We run 256K at $\text{DP}{=}32$, $\text{CP}{=}8$, $\text{GBS}{=}128$ ($4$ GA indices) and 1M at $\text{DP}{=}16$, $\text{CP}{=}16$, $\text{GBS}{=}32$ ($2$ GA indices); \sysname runs at $P{=}8$, $H{=}2$ with $B{=}4096$ (256K) / $8192$ (1M).
The mean straggler reflects throughput without PP, \rev{while the max proxies the PP bottleneck}.
As shown in Figure~\ref{fig:micro_ripple}, \sysname (full) attains the lowest latency on both datasets and both metrics: a mean-straggler speedup over Ulysses of $2.91\times$ (256K) / $2.57\times$ (1M), and a worst-step straggler speedup of $3.14\times$ / $2.90\times$.

\sysname (full) cuts the mean latency over WLB-LLM by $15.9\%$ (256K) / $42.1\%$ (1M), and the max by $68.4\%$ / $65.6\%$: WLB-LLM balances the mean, but its max stays within $0.2\%$ / $0.3\%$ of Ulysses, as indivisible outlier packed sequences still bottleneck its slowest step.
Against DistCA, emulated as a cluster-wide pool ($P{=}\text{DP}{=}32$), \sysname (full) cuts the mean by $53.7\%$ and the max by $64.2\%$ on 256K: DistCA balances globally, but exchanging tiles across all $256$ GPUs inflates its communication.
On 1M ($P{=}\text{DP}{=}16$) DistCA is closer, with \sysname still $7.7\%$ (mean) and $7.9\%$ (max) faster; at larger DP the cluster-wide pool's communication grows further (\S\ref{sec:eval:breakdown}), whereas \sysname keeps $P{=}8$ fixed.

The ablation variants isolate each component's contribution: VRSP trims the max by $9.6\%$ (256K) / $26.2\%$ (1M) over TAP, and Pipeliner overlap trims a further $5.1\%$ / $19.2\%$.
By bounding the pool to $P{=}8$ and rebalancing within it, \sysname avoids both WLB-LLM's worst-step straggler and DistCA's cluster-wide communication cost.

\subsection{\wyRe{Scaling Performance}}
\label{sec:eval:scaling}

We now evaluate \sysname in end-to-end training, where attention-FLOPs imbalance surfaces along two dimensions: across DP ranks it throttles throughput scaling, and across microbatches it inflates pipeline bubbles. \wyRe{Both experiments train the 30B-A3B model (Table~\ref{tab:model_config}).}

\begin{figure}[!t]
    \centering
    \includegraphics[width=\columnwidth]{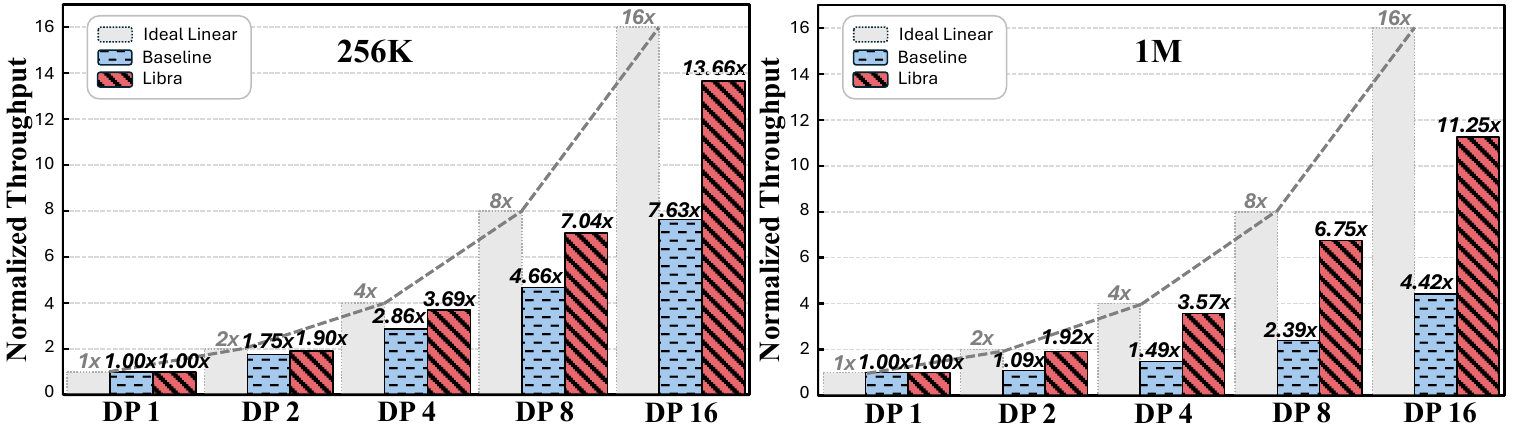}
    \caption{Throughput scaling vs.\ DP size (relative to $\text{DP}{=}1$; ideal is linear) on the 256K and 1M datasets.}
    \label{fig:e2e_dp_scaling}
    \vspace{-5pt}
\end{figure}

\textbf{DP scaling.}
\sysname restores near-linear DP scaling (Figure~\ref{fig:e2e_dp_scaling}).
We fix the global batch size and scale only DP ($\text{PP}{=}1$, $\text{DP}\in\{1,2,4,8,16\}$; 256K: $\text{GBS}{=}128$, $\text{CP}{=}8$, $8$--$128$ GPUs; 1M: $\text{GBS}{=}32$, $\text{CP}{=}16$, $16$--$256$ GPUs), using \sysname (full) with $P{=}\min(8,\mathrm{DP})$ under the same \final{TAP} layout as \S\ref{sec:eval:microbench}, and normalizing throughput to each method's own $\text{DP}{=}1$, where the two coincide (\sysname reduces to $P{=}1$, and Ulysses already balances the single CP group).
As shown in Figure~\ref{fig:e2e_dp_scaling}, at $\text{DP}{=}16$, the baseline reaches only $7.63\times$ (256K) and $4.42\times$ (1M) of the ideal $16\times$ ($47.7\%$ and $27.6\%$ efficiency), as the heaviest DP rank stalls the all-reduce boundary and this penalty grows as larger DP amplifies the long tail.
\sysname reaches $13.66\times$ ($85.4\%$) and $11.25\times$ ($70.3\%$) at the same $\text{DP}{=}16$, an end-to-end speedup of $1.79\times$ and $2.54\times$; the advantage widens with scale and is larger on 1M, whose heavier FLOPs tail leaves more imbalance for global balancing to absorb.

\begin{figure}[h]
    \centering
    \includegraphics[width=\columnwidth]{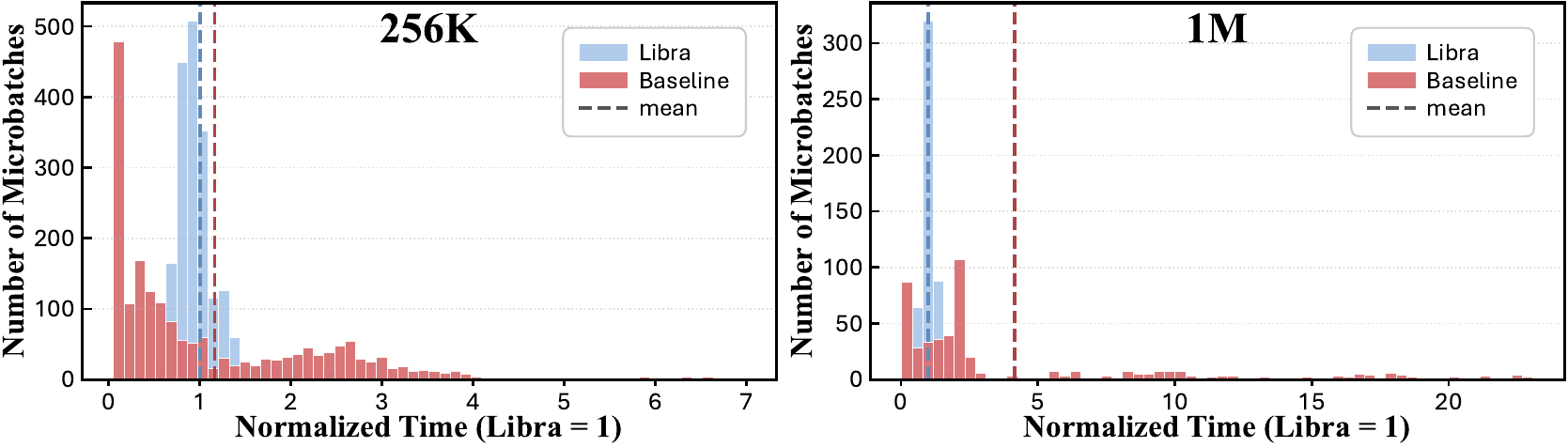}
    \caption{Distribution of per-microbatch forward time on the 256K and 1M datasets. Dashed lines mark the mean.}
    \label{fig:e2e_pp_bubble}
\end{figure}

\textbf{Per-microbatch balance.}
\sysname's global balancing also flattens inter-microbatch imbalance.
We run pipeline-parallel training (256K: $\text{DP}{=}4$, $\text{PP}{=}2$, $\text{CP}{=}8$, $\text{GBS}{=}128$; 1M: $\text{DP}{=}4$, $\text{PP}{=}2$, $\text{CP}{=}16$, $\text{GBS}{=}32$) and measure per-microbatch forward time, aggregated over all ranks of the same stage across $15$ iterations and normalized by \sysname's per-iteration mean (so \sysname's mean sits at $1$).
As shown in Figure~\ref{fig:e2e_pp_bubble}, 
on 1M the baseline spans $0.02$--$23.1$ with a mean of $4.14$, whereas \sysname compresses the range to $0.52$--$1.57$.
Since the pipeline bubble is determined by the slowest microbatch in the window, this worst case is what matters: \sysname cuts it by $14.7\times$ on 1M ($23.1 \to 1.57$) and by $2.6\times$ on 256K ($6.98 \to 2.63$).

\subsection{Pool Size Sweeping and Breakdown}
\label{sec:eval:breakdown}

\begin{figure}[ht]
    \centering
    \includegraphics[width=\columnwidth]{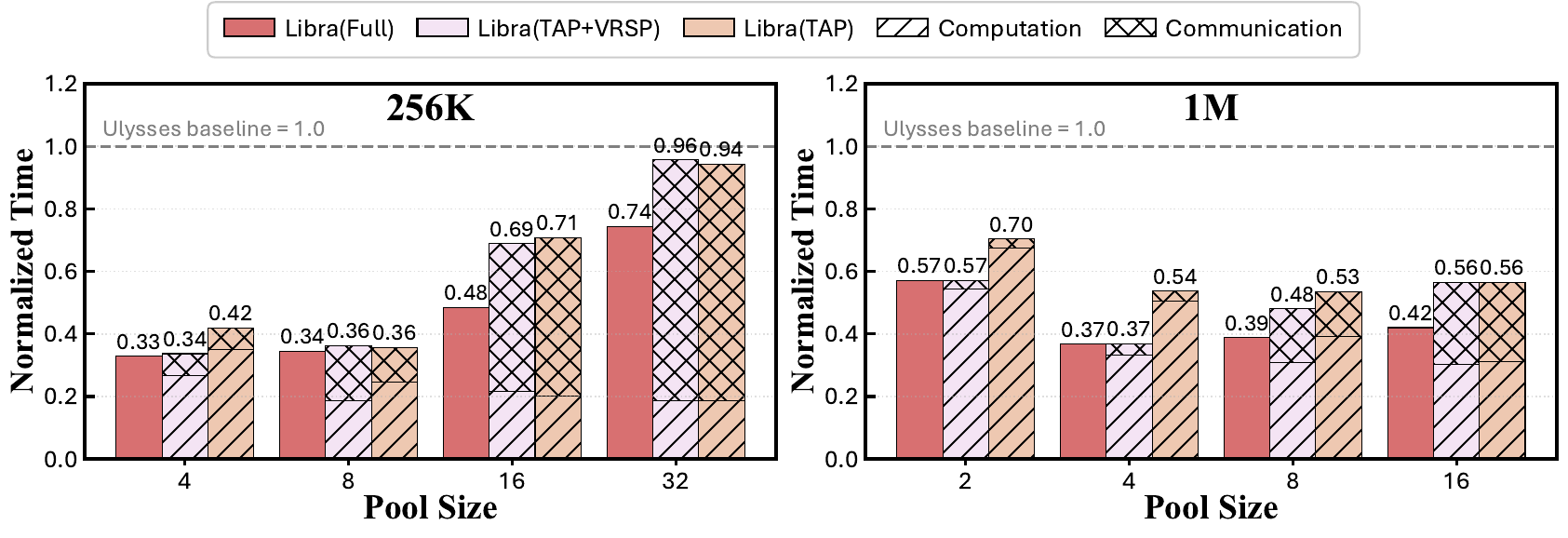}
    \caption{\wy{Per-step core-attention latency vs.\ pool size (256K, 1M), normalized to the Ulysses mean.}}
    \label{fig:micro_breakdown}
\end{figure}

We sweep pool size $P$ (under the same \final{TAP} layout as \S\ref{sec:eval:microbench}) and report the straggler time normalized to Ulysses mean \wyRe{(Figure~\ref{fig:micro_breakdown})}.
\emph{\sysname (TAP} and  \emph{TAP+VRSP)} \compresses{split} latency by profiler item into computation (FlashAttention kernels) and communication (the rest); \emph{\sysname (full)} overlaps \compresses{them}.

\emph{Communication scales with pool size} for two reasons: rebalancing over a wider scope exchanges more tensor volume, and a larger pool pushes a growing share of its traffic onto slow inter-supernode links.
On 256K, communication climbs from $0.07$ at $P{=}4$ to $0.77$ at $P{=}32$; on 1M it stays near $0.03$ through $P{=}4$, then reaches $0.26$ at $P{=}16$, while computation stays flat beyond $P{=}8$ ($0.19$ and $0.31$).
Bounding the pool is thus essential.

\emph{Computation decreases with pool size and saturates at $P{=}8$.}
The slowest pool's computation falls from $0.27$ at $P{=}4$ to $0.19$ at $P{=}8$ on 256K and from $0.54$ at $P{=}2$ to $0.31$ at $P{=}8$ on 1M, with no further gain at larger pools: Figure~\ref{fig:methods_comparison} shows VRSP already eliminates inter-pool imbalance at $P{=}8$ ($R{-}1$ below $0.7\%$), so a larger pool only adds communication overhead.

\subsection{Intra-Pool Grid Searching of TAP}
\label{sec:eval:grid}

\begin{figure}[htbp]
    \centering
    \includegraphics[width=\columnwidth]{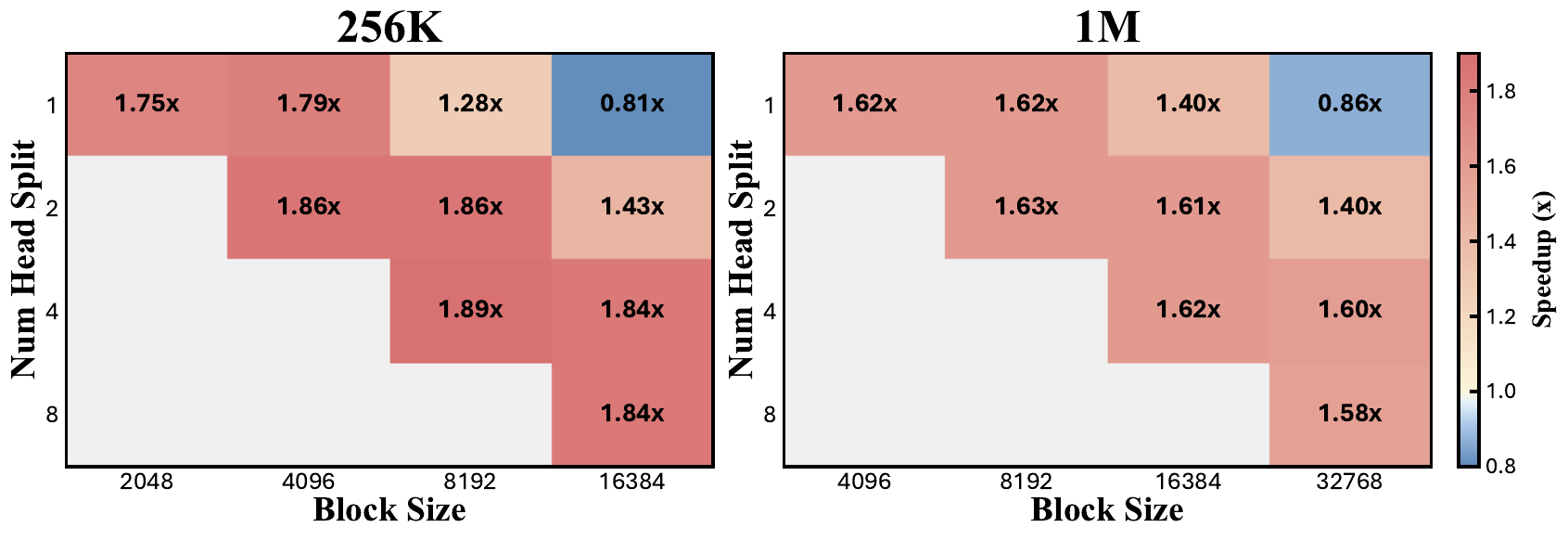}
    \caption{\wy{Grid search of TAP's two granularity knobs.}}
    \label{fig:tap_grid}
    \vspace{-5pt}
\end{figure}

\compresses{TAP's two granularity knobs}, block size $B$ and head split $H$, both add schedulable SH-Tiles and thus improve intra-pool balance, but they differ sharply in communication cost.
We grid-search them on 256K ($\text{CP}{=}8$, $\text{DP}{=}8$, $P{=}8$, $\text{GBS}{=}128$; $B \in \{2048, 4096, 8192, 16384\}$) and 1M ($\text{CP}{=}16$, $\text{DP}{=}4$, $P{=}4$, $\text{GBS}{=}32$; $B \in \{4096, 8192, 16384, 32768\}$), each with $H \in \{1, 2, 4, 8\}$, VRSP enabled and Pipeliner overlap disabled.
To probe \compresses{varied} KV volumes, this sweep alone uses $h_{kv}{=}16$ on 256K and $h_{kv}{=}4$ on 1M.
Figure~\ref{fig:tap_grid} \compresses{gives} the mean-straggler core-attention speedup over same-layout Ulysses.

At an equal tile count, head splitting beats sequence splitting wherever KV traffic matters.
On 256K, $H{=}4, B{=}8192$ and $H{=}1, B{=}2048$ both yield $128$ tiles, yet reach $1.89\times$ versus $1.75\times$: the FLOPs balance is comparable, but the larger block lets each KV-Tile be shared by more query work (\S\ref{sec:design:tap:placer}), while the smaller block fragments KV reuse.
The gap therefore scales with KV volume: it is pronounced on 256K but vanishes on 1M, whose diagonals are nearly flat ($1.62\times$--$1.63\times$ at $256$ tiles), and it reverses once $H{>}h_{kv}$, where head shards fetch replicated KV ($1.58\times$ vs.\ $1.62\times$ at $H{=}8$ on 1M).
Head splitting is thus the preferred way to add granularity, provided the tile count remains sufficient: at $H{=}1$ the 256K speedup collapses from $1.75\times$ ($128$ tiles) to $0.81\times$ ($16$ tiles).

\subsection{Numerical Fidelity and Memory Overhead}
\label{sec:eval:discussion}

\wy{\textbf{Numerical fidelity.}}
\restruct{\sysname reorders samples only within each global batch, leaving each optimizer step's raw-sample multiset unchanged (\S\ref{sec:design:vrsp}); the sole numerical difference is floating-point reassociation (TAP's tile-boundary repartitioning of the online-softmax accumulation, VRSP's reordered reduction), which does not accumulate across steps.  \texttt{\sysid\_attention} meets FlashAttention's own test-suite tolerance~\cite{fa42026}, where the forward and gradient errors against an fp32 reference stay within $2\times$ the bf16 PyTorch reference's errors.}

\wy{\textbf{Memory overhead.}
\sysname adds two transient per-layer memory allocations, including the fetched KV groups (\S\ref{sec:design:tap:exchange}) and the Pipeliner's chunk buffers (\S\ref{sec:design:pipeliner}). 
Both are released at layer end and disjoint from the activations saved for backward; the chunks also reuse a single buffer rather than one per chunk.  End-to-end runs (\S\ref{sec:eval:e2e}) observe at most $5\%$ variation in peak per-GPU memory compared to Ulysses.}

%% file: 7-relatedwork.tex
\section{Related Work}
\label{sec:relatedwork}

\textbf{Attention within a CP group.}
\compresses{Much work} balances attention computation within a context-parallel (CP) group.  Ring schemes include Ring Attention~\cite{ring_attention2023}, Zigzag-Ring~\cite{korthikanti2023reducing}, and Striped Attention~\cite{striped_attention2023}; they circulate KV blocks and arrange sequence shards to reduce causal imbalance.  Burst\-Engine~\cite{BurstEngine2025} overlaps ring communication with attention compute for million-token sequences.  Ulysses~\cite{ulysses2023} exchanges tensors along the head axis\compresses{; LoongTrain}~\cite{loongtrain2024} combines head and sequence parallelism.  Workload-aware dispatchers \compresses{like} MagiAttention~\cite{magiattention2025} and FCP~\cite{fcp2026} place attention tasks within a CP group for heterogeneous masks or workloads.  These systems improve memory scalability, execution, and load balance \emph{within} \compresses{a} CP group.  \sysname instead addresses workload skew \emph{across} CP groups processing different packed sequences, while \compresses{compatible with intra-group attention schemes}.

\textbf{Data repacking and adaptive parallelism.}
\restruct{A complementary line targets global attention imbalance by repacking input data or adapting the parallelism configuration.  LongAlign~\cite{longalign2024} batches by length; WLB-LLM~\cite{wlbllm2025} combines variable-token packing with fine-grained sharding; ChunkFlow~\cite{ChunkFlow2025} merges or splits inputs into uniform chunks; Skrull~\cite{skrull2025} co-optimizes data scheduling with context parallelism.  FlexSP~\cite{flexsp2025}, Dynamic-CP~\cite{dynamiccp2026} and DCP~\cite{dcp2025} vary the sequence-parallel or CP degree per sequence or sequence bucket.  Zeppelin~\cite{zeppelin2026} instead migrates \compresses{data} at runtime, placing length-zoned chunks topology-aware across DP ranks and remapping surplus tokens \compresses{for} linear layers.
These approaches cannot balance a CP group holding an indivisible outlier (\S\ref{sec:motivation:scope}), and their per-microbatch granularity (a CP degree or a length bucket) is too coarse for the hundreds-fold compute gap between 1M- and 32K-token samples.  \sysname instead retains fixed-token packing and fixed pool membership, redistributing parameter-free core-attention tasks within a bounded sequence pool while preserving the optimizer-step raw-sample multiset and the parallel schedule.}

\textbf{Core-attention disaggregation.}
\restruct{DistCA~\cite{DistCA2025} disaggregates core attention from model workers onto a dedicated attention-server pool that dynamically rebatches token-level tasks for global balance.  This cluster-wide design maximizes scheduling scope but breaks the standard pipeline schedule and ties task movement to slow cross-machine bandwidth.  MoE models leave its ping-pong cross-layer overlap little room: expert routing and all-to-all exchange fragment MLP phases.  \sysname instead keeps the redistribution domain bounded---each sequence pool stays within one PP stage and spans $P$ DP replicas---so DP scale-out adds fixed-size pools rather than enlarging each domain.}

%% file: 9-conclusion.tex
\section{Conclusion}
\label{sec:conclusion}

Through \sysname, we demonstrate the effectiveness of applying the law of large numbers to attention load balancing in long-tailed, long-context LLM training.
Specifically, \sysname presents three innovations to address attention FLOPs imbalance: 
1)~using the law of large numbers as a scaling principle and forming the attention pools, 
2)~introducing VRSP to accelerate inter-pool balance, 
and 3)~developing TAP to balance 2D SH-Tiles within each pool, and overlapping the resulting tensor movement with attention computation.
And \sysname delivers up to $2.54\times$ end-to-end throughput on Qwen3-30B-A3B training.
Going forward, we hope \sysname draws attention to bounded, pool-level statistical balancing as a practical approach to long-tailed workloads in distributed training.